\documentclass[a4paper]{article}
\usepackage[ansinew]{inputenc}
\usepackage{times}
\usepackage[T1]{fontenc}
\usepackage{graphicx}
\usepackage{geometry}
\usepackage{amssymb}
\usepackage{amsmath}

\usepackage{rotating}

\usepackage{tcolorbox}
\usepackage{xcolor}
\colorlet{shadecolor}{gray!25}
\usepackage{subfig}
\usepackage{float}

\author{\normalsize Ludwig A. Hothorn,\\ 
\footnotesize Im Grund 12, D-31867 Lauenau, Germany (e-mail:ludwig@hothorn.de)\\ \scriptsize(retired from Leibniz University Hannover}

\title{Simultaneous comparisons of the variances of k treatments with that of a control: a Levene-Dunnett type procedure}
\begin{document}

\maketitle

\begin{abstract}
There are some global tests for heterogeneity of variance in k-sample one-way layouts, but few consider pairwise comparisons between treatment levels.
For experimental designs with a control, comparisons of the variances between the treatment levels and the control are of interest - in analogy to the location parameter with the Dunnett (1955) procedure. Such a many-to-one approach for variances is proposed using the Levene transformation, a kind of residuals. Its properties are characterized with simulation studies and corresponding data examples are evaluated with R code.
\end{abstract}


\section{A  Levene-Dunnett type test}
Several tests (or pre-tests) on variance homogeneity in completely randomized one-way layouts exists for independent pairwise ratios, for global tests \cite{Katsileros2024} or for ordered alternatives \cite{Pallmann2014}. In clinical dose-finding studies and toxicological bioassays, simultaneous comparisons of the variances of $k$ treatments with that of a control are of interest in a design with treatment levels  $T_j, j=0,...,k$, i.e. $[T_j: T_0,T_1, T_2, ..., T_k]$ is commonly used, where $T_0$ represents a placebo or a negative control group. In principle the closed testing procedure can be used to achieve simultaneous, pairwise comparisons against control \cite{Boos2023}. Whereas adjusted p-values are easy to estimated, simultaneous confidence intervals are complicated \cite{Hothorn2020}. A straightforward approach is the use of Levene transformed variables $Z_{ji}=Y_{ji}-Median(Y_j)$ directly within the standard Dunnett procedure \cite{Dunnett1955}. The common Dunnett test \cite{Dunnett1955} can be formulated as a multiple contrast test: $maxT=max(t_1,...,t_k)$ where $t_\xi$ are standardised contrast tests: $ t_\xi=\sum_{i=0}^k c_i\bar{y}_i/S \sqrt{\sum_i^k c_i^2/n_i}$ where $c_i^\xi$ are the contrast coefficients. For a balanced design with $k=2$ the contrast matrix is simple, e.g. for the one-tailed Dunnett test:

\begin{table}[ht]
\centering\small
				\begin{tabular}{ c  c c r c c c c c}
         $c_i$ & C & $T_1$ & $T_2$ & $T_2$\\ \hline
        $c_a$ & -1 & 0  &0   & 1   \\
        $c_b$ & -1 & 0  & 1 & 0  \\
				 $c_b$ & -1 & 1 & 0 &0  \\
				\end{tabular}
				\caption{Dunnett-type contrast matrix for $k=1+3$ comparisons}
        \end{table}

The quantiles of the max-T test $t_{k,df,R,1-\alpha}$ are monotonic with $k$ which provides elementary inference, i.e. adjusted p-values and/or simultaneous confidence intervals. As in the Dunnett original procedure, one- or two-sided hypothesis formulations are easily possible.\\
The empirical behavior of familywise error rate (FWER) and power of this new Levene-Dunnett test (LevDun) is characterized by means of a simulation study, whereby the behavior with the usual small $n_i$ is of interest. The estimation of both adjusted p-values and simultaneous confidence intervals is demonstrated using an example.

\section{Simulations}
\subsection{Simulations under global $H_0$}

First,  a balanced design $k=3+1$ is considered.
\begin{table}[ht]
\centering
\scalebox{0.75}{
\begin{tabular}{rr}
  \hline
 $n_i$ & $\hat{\alpha}$  \\ 
  \hline
 3 & 0.001  \\ 
 5 & 0.010  \\ 
 10 & 0.042  \\ 
 20 &  0.044 \\ 
 30 &  0.046  \\ 
 50 &  0.045  \\ 
   \hline
\end{tabular}
}
\caption{Empirical FWER $\hat{\alpha}$ in balanced design ($k=3+1, n_i=const$, one-sided)}
\end{table}

This test, based on an asymptotic transformation, controls the FWER from about $n_i=10$ on.

Second, selected unbalanced designs for $k=3+1$ are considered.

\begin{table}[ht]
\centering
\scalebox{0.75}{
\begin{tabular}{rr}
  \hline
 $n_i$ & $\hat{\alpha}$  \\ 
  \hline
 10,10,10,10 & 0.042  \\ 
 16,8,8,8,8 & 0.033  \\ 
 8,8,8,12 & 0.041  \\
   \hline
\end{tabular}
}
\caption{Empirical FWER $\hat{\alpha}$ in unbalanced designs ($k=3+1, N=40$, one-sided)}
\end{table}
For selected unbalanced designs, the test becomes slightly conservative
\subsection{Simulations under partial $H_1$}

Different patterns of variance heterogeneity are considered in a balanced design.

\begin{table}[ht]
\centering
\scalebox{0.85}{
\begin{tabular}{rrrr|rrrr}
  \hline
$\sigma_0$ & $\sigma_1$  & $\sigma_2$ & $\sigma_3$  & $\hat{\pi}$ & $\hat{\pi/\alpha_{01}}$ & $\hat{\pi/\alpha_{02}}$  & $\hat{\pi/\alpha_{03}}$  \\ 
  \hline
 const & const & const & $\Uparrow$ & 0.806 & 0.002 & 0.000 & 0.806 \\ 
 const & const & $\Uparrow$ & $\Uparrow$ & 0.847 & 0.000 & 0.617 & 0.598 \\ 
 const & $\Uparrow$ & $\Uparrow$& $\Uparrow$ & 0.789 & 0.417 & 0.403 & 0.413 \\ 
 const & $\Uparrow$ & $\Uparrow$ & const& 0.849 & 0.598 & 0.597 & 0.000 \\ 
const & $\Uparrow$ & const& const & 0.822 & 0.820 & 0.004 & 0.001 \\ 

 const & const & $\Uparrow$ & const & 0.810 & 0.001 & 0.810 & 0.003 \\ 
 const & $\Uparrow$ & const & $\Uparrow$ & 0.836 & 0.593 & 0.000 & 0.584 \\ \hline
 $\Uparrow$ & const & const & $\Uparrow$ & 0.049 & 0.000 & 0.000 & 0.049 \\ \hline
    \hline
\end{tabular}
}
\caption{Global and elementary powers (size) for various variance patterns ($n_i=10$, one-sided)}
\end{table}
The power estimates depend on the particular variance pattern, the elementary power estimates are correct with respect to the underlying pattern, as are the elementary $\alpha_i$. The one-sided version provides a correct directional testing.

Patterns of variance heterogeneity are also considered for selected unbalanced designs.
\begin{table}[ht]
\centering
\scalebox{0.85}{
\begin{tabular}{rrrr|rrrr}
  \hline
$n_0$ & $n_1$  & $n_2$ & $n_3$  & $\hat{\pi}$ & $\hat{\pi/\alpha_{01}}$ & $\hat{\pi/\alpha_{02}}$  & $\hat{\pi/\alpha_{03}}$  \\ 
  \hline
 10 & 10 & 10 & 10 & 0.806 & 0.002 & 0.000 & 0.806 \\ 
 16 & 8 & 8 & 8 & 0.799 & 0.002 & 0.799 & 0.001 \\ 
 16 & 10 & 4 & 10 & 0.607 & 0.010 & 0.599 & 0.006 \\ 
   \hline
\end{tabular}
}
\caption{Global and elementary powers (size) for a selected variance pattern ($\sigma_0 =\sigma_1< \sigma_2 \sigma_3$) in unbalanced design ($N=40$, one-sided)}
\end{table}

Hardly any different power for a design with a higher sample sizes in the control $n_0$ (i.e., an optimal design for location effects), but, as expected, a significant drop in power when the group with the increased variance uses a significantly lower sample size (still for a constant total sample size N).

\section{A modified Levene transformation}
For designs with small sample sizes $n_i$, in treatment groups with odd $n_i$ the value $Z_{ji}=0$ occur which cause conservativeness \cite{Boos2023}. A one-sided trimmed test version, omitting this single value zero, (modLevDun) avoid this. 

\begin{table}[ht]
\centering
\scalebox{0.75}{
\begin{tabular}{rrr}
  \hline
 $n_i$ & LevDun & modLevDun \\ 
  \hline
15 & 0.036 &0.051 \\ 
11 &  0.029 &0.048  \\ 
7 &  0.019 & 0.050  \\ 
5 &  0.010 & 0.040  \\ 
3 &  0.001 & 0.047 \\ 
   \hline
\end{tabular}
}
\caption{Small sample size behavior: comparison of FWER estimates of Levene and  modified Levene-Dunnett test}
\end{table}

\begin{table}[ht]
\centering
\scalebox{0.65}{
\begin{tabular}{rrrr|rrrr|rrrr}
  \hline
 $\sigma_1$ & $\sigma_2$ & $\sigma_3$ & $\sigma_4$ & LevDun & $LevDun_{01}$ & $LevDun_{02}$ & $LevDun_{03}$ & modLevDun & $modLevDun_{01}$ & $modLevDun_{02}$ & $modLevDun_{03}$ \\ 
  \hline
 const &  const &  const & $\Uparrow$ & 0.845 & 0.001 & 0.000 & 0.845 & 0.869 & 0.001 & 0.000 & 0.869 \\ 
  const &  const & $\Uparrow$ &  const & 0.867 & 0.000 & 0.866 & 0.001 & 0.883 & 0.000 & 0.882 & 0.001 \\ 
  const & $\Uparrow$ &  const &  const & 0.848 & 0.848 & 0.000 & 0.000 & 0.878 & 0.878 & 0.001 & 0.000 \\ 
 $\Uparrow$ &  const &  const &  const & 0.000 & 0.000 & 0.000 & 0.000 & 0.000 & 0.000 & 0.000 & 0.000 \\ 
  const &  const & $\Uparrow$ & $\Uparrow$ & 0.879 & 0.000 & 0.620 & 0.637 & 0.911 & 0.000 & 0.688 & 0.698 \\ 
 const & $\Uparrow$& $\Uparrow$& $\Uparrow$ & 0.861 & 0.444 & 0.452 & 0.451 & 0.910 & 0.521 & 0.531 & 0.541 \\ 
   \hline
\end{tabular}
}
\caption{Comparison of elementary power estimates of Levene and  modified Levene-Dunnett test for balanced small sample size $n_i=11$}
\end{table}

The control of the FWER even at small $n_i$ and thus the higher power are impressive, although this modification uses a biased test statistic.

\section{A global test for any heterogeneity}
Sometimes global test results are of interest in unstructured k-sample designs. Instead the Dunnett many-to-one comparison contrasts, the multiple contrasts against the grand mean can be used \cite{Konietschke2013}. While conventional F-tests use the sum of the quadratic deviation from the overall mean, the maximum test using linear deviations is used here. Such an approach allows both global claims (using the $min(p_x)$ approach) and adjusted p-values or simultaneous confidence intervals for the pairwise comparisons against the grand mean. This approach is demonstrated by an example below.

\section{Examples}
The first example uses the litter weight data in a toxicological bioassay (data available in library(multcomp)) \cite{Westfall1997}. The R-code is simple to estimate adjusted p-value or simultaneous confidence intervals
\small
\begin{verbatim}
library(multcomp)
data(litter)
medrs<-tapply(litter$weight, litter$dose, median) 			# group-specific medians
medrsV<-medrs[as.integer(litter$dose)] 									# median for each subject
litter$tresponse<- abs(litter$weight-medrsV) 						# Levene transformation
mod2<-lm(tresponse~dose, data=litter)										# linear model	
summary(glht(mod2, linfct = mcp(dose ="Dunnett"), alternative="greater"))
plot(glht(mod2, linfct = mcp(dose ="Dunnett"), alternative="greater"))
\end{verbatim}
\normalsize

\begin{figure}[ht]
	\centering
		\includegraphics[width=0.40\textwidth]{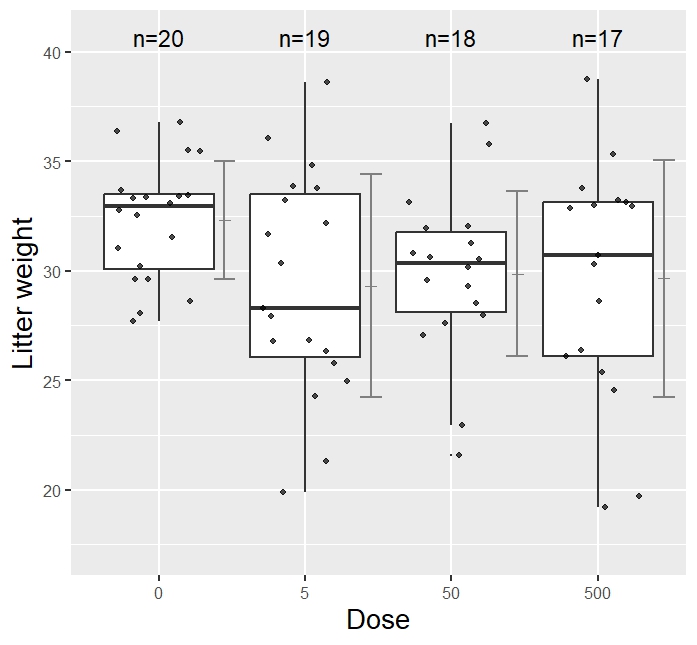}
	\caption{Boxplot of litter weight data}
	\label{fig:west2024}
\end{figure}

\begin{table}[ht]
\centering
\begin{tabular}{l|r|rr||r}
  \hline
 Comparisons & p-value LevDun   \\ 
  \hline
 5 - 0  & 0.022  \\ 
 50 - 0 & 0.491   \\ 
500 - 0 & 0.022 \\ 
   \hline
\end{tabular}
\caption{p-values for one-sided Levene-Dunnett test for the litter weight data}
\end{table}

\begin{figure}[ht]
	\centering
		\includegraphics[width=0.4\textwidth]{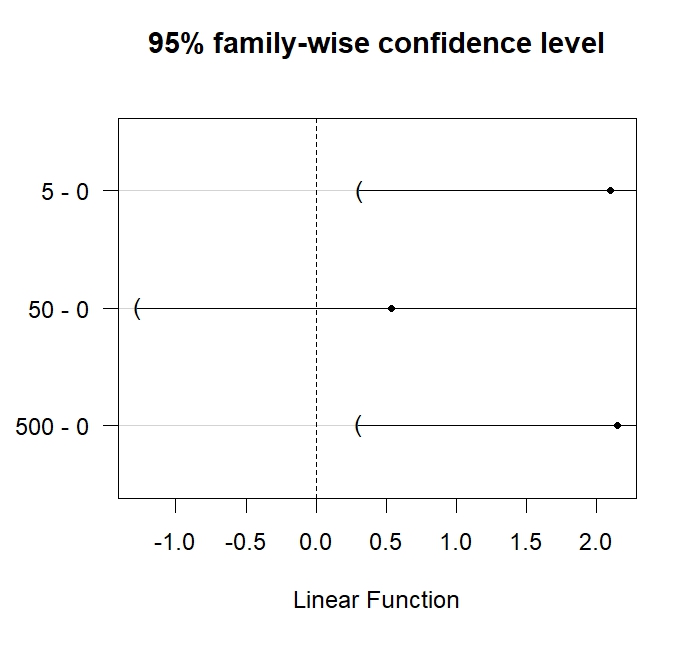}
	\caption{Simultaneous confidence intervals for Levene transformations}
	\label{fig:westfallVarCI2024}
\end{figure}

The second example uses the survival time data in \cite{Boos2023} as an example for unstructured k-sample design:
\begin{figure}[h]
	\centering
		\includegraphics[width=0.450\textwidth]{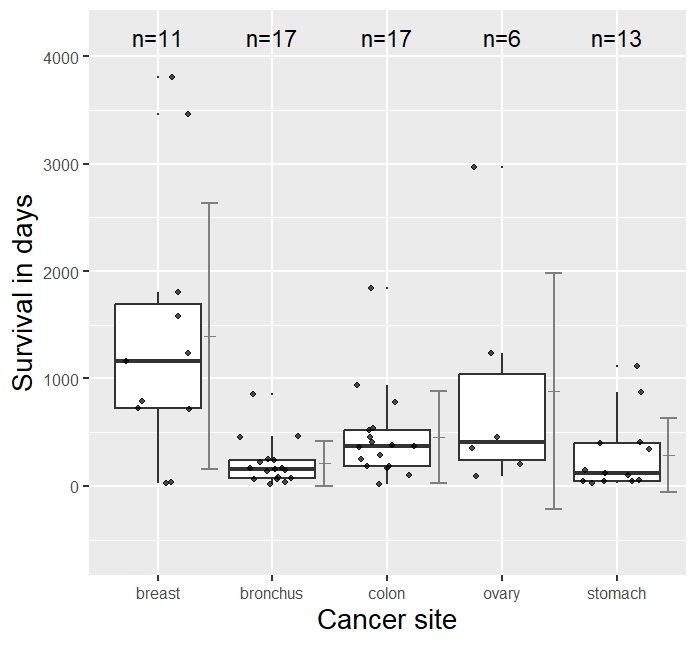}
	\caption{Boxplot for Boos et al. 2023 example}
	\label{fig:BoxBoos}
\end{figure}

\small
\begin{verbatim}
boos<-data.frame(survival, group=as.factor(grp))
library(multcomp)
medrs<-tapply(boos$survival, boos$group, median) # group-specific medians
medrsV<-medrs[(boos$group)] # median for each subject
boos$tresponse<- abs(boos$survival-medrsV) # Levene transformation
mod3<-lm(tresponse~group, data=boos)
summary(glht(mod3, linfct = mcp(group="GrandMean")))
library(GLDEX)
no0<-fun.zero.omit(boos$tresponse)
boos0<-data.frame(no0, group0=as.factor(grp0))
mod4<-lm(no0~group0, data=boos0)
summary(glht(mod4, linfct = mcp(group0="GrandMean")))
\end{verbatim}
\normalsize

\begin{table}[ht]
\centering
\begin{tabular}{l||l|l|}
  \hline
 Cancer site & LevDun & modLevDun   \\ 
  \hline
 breast & 0.002  & 0.002\\ 
 bronchus & 0.999 & 0.13  \\ 
colon & 0.999 &0.78 \\
ovary & 0.31& 0.54 \\
stomach & 0.99& 0.79 \\ \hline
global Levene min(p)-test &0.002 & 0.002\\
global Levene F-test &0.027 & -\\
   \hline
\end{tabular}
\caption{p-values for 2-sided Levene-Dunnett and modLevDun test for survival data}
\end{table}

The advantage of the new heterogeneity test compared to the F-test variant (in library(misty)) is obvious: not only is a global p-value is available (i.e. 'variance heterogeneity exists'), but the statement 'the variance at the tumor site='breast' is increased with respect to the overall variance' is also available.

\section{Summary}
The proposed Levene-Dunnett test can be recommended for randomized designs with a control or placebo group for the analysis of possible variance heterogeneity, whereby the modified version should be used for small sample sizes. An analogous heterogeneity test for unstructured designs is also possible. Two data examples were used to illustrate the simple R code and the problem-adapted interpretation.


\end{document}